  \documentstyle{elsart}
  
  \def\n{\nonumber}
  
  \def\th{\theta}

  \def\be{\begin{equation}}
  \def\ee{\end{equation}}
  \def\bq{\begin{eqnarray}}
  \def\eq{\end{eqnarray}}
  
  \def\({\left(}
  \def\){\right)}

\begin{document}
  \begin{frontmatter}
  \title{Exact solutions for null fluid collapse in higher dimensions}
  \author{L.K. Patel$^1$} and
  \author{Naresh Dadhich$^2$},
 
 {$^1$Department of Mathematics, Gujarat University, Ahmedabad 380009,
India}

 {$^2$Inter--University Center for Astronomy and Astrophysics, 
  Post
  Bag 4, Ganeshkind, Pune 411007, India}

 {E-mail: naresh@iucaa.ernet.in}
  
  \begin{abstract}
  A large family of inhomogeneous non-static spherically symmetric 
solutions of the Einstein equation for null fluid in higher dimensions 
has been obtained. It encompasses higher dimensional versions of many 
previously known solutions such as Vaidya, charged Vaidya and Husain solutions
and also some new solutions representing global monopole or string dust. 
It turns out that physical properties of the solutions carry over to 
higher dimensions.
\end{abstract}
  
  \n PACS numbers: 04.20Jb, 04.40Nr, 04.40+c
 
  
  \end{frontmatter}
  
  \section{Introduction}
 
 Inspired by work in string theory and other field theories, there has 
been a considerable interest in recent times to find solutions of the 
Einstein equation in dimensions greater than four. It is believed that 
underlying spacetime in the large energy limit of the Planck energy may have 
higher dimensions than the usual four. At this level, all the basic 
forces of Nature are supposed to unify and hence it would be pertinent in 
this context to consider solutions of the gravitational field equation in 
higher dimensions. Of course this consideration would be relevant when 
the usual four dimensional manifold picture of spacetime becomes  
inapplicable. This would happen as we approach singularity whether in 
cosmology or in gravitational collapse.

 Chodos and Detweiler [1] have considered the Kasner vacuum solution in 
5-dimansional Kaluza-Klein (K-K) theory and the question of dimensional 
reduction. There have been investigations of entropy and classification of 
homogeneous cosmologies in K-K theory [2-3]. For field of localized 
sources, higher dimensional versions of Schwarzschild, 
Reissner-Nordstr$\ddot {\rm o}$m, Vaidya [4-5] and Bonnor-Vaidya [6] have been 
considered by several authors [7-10]. Important issues like formation of 
primordial black holes have also been addressed to [8]. 

 Gravitational collapse continues to occupy centre-stage in gravitational 
research ever since mid sixties. The question is  what initial 
conditions lead to formation of black hole or naked singularity. There 
has been some interesting investigations in this direction which has been 
reviwed in [11]. The use of Vaidya solution was first made by Papapetrou 
[12] in studying collapse and he pointed out that it may lead to 
formation of naked singularity. Joshi et al [13] have done an 
extensive analysis of gravitational collapse using the Vaidya metric in 
the context of naked singularity. It is an important question, whether 
collapse always leads to singularity hidden behind a black hole event 
horizon or it is naked [14]? Vaidya [4] and Bonnor-Vaidya [6] solutions have 
been employed to study various aspects of black hole formation [15-16]. 
Recently Husain [17] has considered collapse of a Type II null fluid 
with an equation of state and has obtained some non-static spherically 
symmetric exact solutions of the Einstein equation. It turns out that 
the metrics representing null fluid collapse have multiple apparent  
horizons and in the large  limit asymptotically flat metrics have 
short hair and they can be thought of "lying between" Schwarzschild and 
Reissner-Nordstr$\ddot {\rm o}$m metrics [18]. Wang [19] has very recently 
obtained a large family of inhomogeneous and non-static spherically symmetric 
solutions that includes all the previously known solutions. In this paper 
we wish to obtain the higher dimensional version of this large family and 
then consider a particular solution to show that all the collapse 
properties discussed by Husain carry over to higher dimensions. Further 
we also obtain solutions for a black hole with a global monopole charge 
[20-21] or sitting in a string dust [22] universe. It should be noted 
that this synthesis of solutions is made possible by the fact 
that energy-momentum tensor is linear in mass functions.

 In section 2, we obtain the generalized Vaidya family in higher 
dimensions which would be followed in section 3 by discussion of various 
particular cases synthesized in the family. Finally we shall conclude 
with discussion of collapse properties of an analogue of the Husain 
solution. 

\section{Generalized Vaidya family}

 Let us consider $(n + 2)$-dimensional spherical spacetime described by 
the metric,
\bq   
ds^2 = 2 d u d r + (1 - \frac{2 m(u,r)}{(n-1)r^{n-1}})d u^2 - 
r^2 d {\omega}_n^2 \label{1}
\eq
\n where

\bq
d {\omega}_n^2 = \d\th_1^2 + sin^2\th_1\d\th_2^2 + ... + 
sin^2 \th_{n-1} d \th_n^2 \label{2} 
\eq

\n is the metric on $n$-space in spherical polar coordinates. Here $m$ is 
an arbitrary function of retarded time $u$ and radial coordinate $r$. 
When $m = m(u)$, it is the Vaidya solution in higher dimensions [9]. The 
usual Vaidya solution in 4-spacetime follows for $n=2$.

 Let us name the coordinates as follows:
\bq
x^0 = u, ~x^1 = r, ~x^i+1 = \th_i, ~i = 1,2,...,n. \label {3}\eq
\n In the standard way we compute the Einstein tensor and it reads as

\bq
G_{00} =  \frac{n \dot m} {(n-1)r^n} - \frac{(n m ^{\prime} }  {(n-1)r^n}[1 - \frac {2m} {(n-1)r^{n-1}} ]\cr
 G_{01} = - \frac{n  m ^{\prime} }{(n-1)r^n},\cr
G_{22} = \frac{ m ^{\prime \prime}}{(n-1)r^{n-3}},\cr
G_2^2 = G_3^3 = ... = G_{n+1}^{n+1}. \label{4} 
\eq

\n Here and in what follows an overhead dash and dot denote derivative 
relative to $r$ and $u$ respectively.

 The energy-momentum tensor for a Type II fluid is given by [23,18],
\bq
T_{ik} = \mu l_il_k +(\rho + p)(l_i\eta_k + l_k\eta_i) - p g_{ik} 
\label{5}\eq
\n where
\bq
l_il^i = \eta_i\eta^i = 0, ~l_i\eta^i = 1. \label{6} \eq
\n The null vector $l_i$ is a double null eigenvector of $T_{ik}$. 
Physically occurring distribution is null radiation flowing in the radial 
direction correpsponding to $\rho = p = 0$, the Vaidya spacetime of 
radiating star. When $\mu = 0$, $T_{ik}$ reduces to degenerate Type I 
fluid [23] and further it represents string dust for $\mu = 0 = p$.

 The energy condition for such a distribution are as follows [23]: 

(i) the weak and strong energy conditions,
\bq
\mu > 0, ~\rho \geq 0, ~p \geq 0 \label{7} 
\eq

(ii) the dominant energy condition,
\bq
\mu > 0 ~\rho \geq p \geq 0 .\label{8} 
\eq

\n In the case of $\mu = 0$, the energy conditions would become,

(iii) the weak condition
\bq
\rho + p \geq 0, ~\rho \geq 0 \label{9} 
\eq

\n (iv) the strong condition

\bq
\rho + p \geq 0, ~p \geq 0 \label{10} 
\eq

\n (v) the dominant

\bq
\rho \geq 0 ~-\rho \leq p \leq \rho. \label {11} 
\eq

\n The energy-momentum tensor (\ref{4}) has support along both the two 
future pointing null vectors $l_i$ and $\eta_i$, and it is exactly, as we 
shall show later, in the form to give Bonnor-Vaidya metric in higher 
diemnsions. We also note that $T_{ik}l^il^k = 0$ and $T_{ik}\eta^i\eta^k 
= \mu$.

 For the metric (\ref{1}) we write
\bq
l_i = g_i^0, ~\eta_i = g_i^1 + \frac{1}{2}(1 - 
\frac{2m}{(n-1)r^{n-1}})g_i^0. \label{12} 
\eq

\n We wish to solve the Einstein equation

\bq
G_{ik} = -8 \pi T_{ik}. \label{13} 
\eq

\n Note that (\ref{11}) satisfies the conditions (\ref{5}).

Substituting (\ref{4}) in (\ref{12}) we obtain

\bq
8\pi\mu = -\frac{n \dot m}{(n-1)r^n}, ~8 \pi \rho = \frac{n  m ^{\prime } }{(n-1)r^n}, ~8 \pi p = -\frac{ m ^{\prime \prime}}{(n-1)r^{n-1}}. \label{14} 
\eq

 \n The part $\mu l_il_k$ of $T_{ik}$ in (\ref{4}) is the component of 
matter field that moves along the null hypersurface $u = const.$. In 
particular when $p = \rho = 0$ we have Vaidya solution in higher 
dimensions. Thus the distribution in (\ref{13}) represents Vaidya 
radiating star in Type II fluid universe in higher dimensions. Note that 
when $\mu = 0$, though we have $\rho, p$ given by (\ref{13}), $T_{ik}$ does not 
reduce to the perfect fluid form. It would rather represent an imperfect 
fluid.

 By proper choice of the mass function, the energy conditions can be 
satisfied. Without loss of generality we write

\bq
m(u,r) =  \Sigma^{-\infty}_ {\infty} a_i(u)r^i \label{15} 
\eq

\n where $a_i(u)$ are arbitrary functions of the retarded time $u$. The 
summation would go over to integral for continuous summation index. 
Substitution of (\ref{14}) into (\ref{13}) yields
\begin{eqnarray}
8\pi\mu = -\frac{n}{n-1}  \Sigma^{-\infty}_ {\infty}  a_ir^{i-n},\cr
8\pi\rho = \frac{n}{n-1} \Sigma^{-\infty}_ {\infty} ia_ir^{i-n-1}  \cr 
8\pi p = -\frac{1}{n-1}  \Sigma^{-\infty}_ {\infty} i(i-1)a_ir^{i-n-1} \label{16} 
\end{eqnarray}

 \n This goes over to Wang family [19] for the 4-dimensional spacetime 
when $n = 2$. Thus ours is the higher dimensional generalization. We 
recover Schwarzschild solution in higher dimensions for $\rho = \mu = p = 
0$. Also note from (ref{14}) that $\rho = 0$ implies $p = 0$. That means 
Vaidya solution  simply follows from $\rho = 0$. This family 
includes many previously known higher dimensional solutions, which we 
consider in the next section.

\section{Particular cases}

 In the following we shall consider some of the particular cases.

(a) Let us choose the functions $a_i(u)$ as

\begin{equation}
 a_i(u) = 
\cases{a/2,&  $i = 1 $ \cr 0,&    $i \neq  1$ }
\label{17}
\end{equation}

\n where $a$ is an arbitrary constant. In this case we shall have

\bq
m(u,r) = \frac{a}{2}r, \ \ \ ~\mu = p = 0, \ \ \ ~8\pi\rho = \frac{na}{2(n-1)r^n}.
\label{18} 
\eq

\n Clearly the matter field is of Type I and satisfies all the energy 
conditions [8-9] for $a >  0$. The metric would read as (without the angular
 part )
\bq
 d s^2 = 2 d u d r + (1 - \frac{a}{r^{n-2}})d u^2 
\label{19} 
\eq

\n which can be identified as higher dimensional description of field of 
a Schwarzschild particle with global monopole [20,21] or of a particle 
sitting in a string dust [22] universe. For $n =2 $, it reduces to the 
monopole solution [20]. 

This is the case which is not considered by Wang [19].

(b) In this case we set
\bq
a_i = \cases{ m_0, & $i = n-1$ \cr
      0,  & $i \neq n-1 $} \label{20} \eq
\n where $m_0$ is a constant. For this case, we have

\bq
m = m_0 r^{n-1}, ~\mu = 0, ~8\pi\rho = nm_0/r^2, ~8 \pi p = -m_0(n-2)/r^2 
\label{21} 
\eq

\n Here again it is Type I distribution satisfying the weak and dominant 
but not strong energy conditions for $m_0 > 0$. It violates the strong 
energy condition because $n \geq 2$. The metric would be given by

\bq
ds^2 = 2d u d r + (1 - 2m_0/(n-1))du^2  \label{22} 
\eq

\n which would also give the global monopole spacetime as above for $n = 2$.

\def\gne #1 #2{\ \vphantom{S}^{\raise-0.5pt\hbox{$\scriptstyle #1$}}_
{\raise0.5pt \hbox{$\scriptstyle #2$}}}

\def\ooo #1 #2{\vphantom{S}^{\raise-0.5pt\hbox{$\scriptstyle #1$}}_
{\raise0.5pt \hbox{$\scriptstyle #2$}}}

(c) For
\bq
a_i =  \cases{ \frac {\wedge (n-1)} {n (n+1)}, &$i = n+1$ \cr
0, &$i \not= n+1 $}\label{23}
\eq
\n we find
\bq               
m(u,r) = \frac {\wedge (n-1)} {n (n+1)} r^{n+1}, 8 \pi p = -8 \pi p = \wedge, \mu = 0 \label {24}
\eq
\n where $\bigwedge$ is the cosmological constant. The metric would read as
\bq
   ds^2 = 2 dudr + ( 1 - \frac {2 \wedge} {2 (n+1)} r^2 )du^2 \label {25}                 
\eq
\n This is the de Sitter and anti de Sitter universe in higher dimensions 
for $\bigwedge g  {\gne > <} ~~0$ and reducing to the familiar de Sitter metric when $n 
= 2$.

(d) We now obtain higher dimensional Bonnor-Vaidya solution of a 
radiating charged particle. For this we choose
\bq
 a_i(u) = \cases{ f(u), &$i = 0$\cr
\frac {-4\pi e^2(u)} {n}, &$ i = l -n $\cr
0, &$i \not= 0, l-n$ }. \label{26}
\eq
\n The two arbitrary functions $f(u)$ and $e(u)$ represent mass and 
electric charge at the retarded time $u$. the physical parameters would 
be given by
 \bq
  m = f(u) - \frac {4 \pi e^2 (u)} {n r^{n-1}}\cr
8 \pi \rho = 8 \pi p = \frac {4 \pi e^2(u)} { nr^2} \cr
8 \pi \mu = - \frac {1} {(n-1) r^n} ( n \dot f - \frac {8 \pi e \dot e} {r^{n-1}} )\label{27}
 \eq

 \n Clearly $\rho, p$ are always positive, the only condition that would 
restrict the functions $f(u)$ and $e(u)$ $\mu \geq 0$. If $\dot f > 0$ 
and $\dot e < 0$, then the energy conditions (7-8) will be 
satisfied. The metric would be given by
\bq
 ds^2 = 2dudr  + [1 - \frac {2 f(u)} {(n-1)r^{n-1}} + \frac {8 \pi e^2(u)} {n(n-1)r^{2(n-1)}} du^2  \label{28}
\eq

 \n This is the Bonnor-Vaidya solution in higher dimensions [10]. The 
electromagnetic field is given by
\bq
 F_{ik} = \frac {e(u)} {r^n} \left( g^{\prime}_i g^0_k - g^0_i g^{\prime}_k\right) \label{29}
\eq
\n with the 4-current vector,
\bq
  4 \pi J^i = - \frac {\dot e (u)} {r^n} g^i_1.\label{30}
\eq

 (e) Finally we consider the higher dimensional analogue of Husain 
solution [17], which would require
\bq
 a_i(u) = \cases{ f(u), &$i = 0$ \cr
- \frac {g(u)} {nk-1}, &$i =1 - nk, (k \not= l/n)$\cr
0, &$i \not= 0, 1 -nk$}\label{31}
\eq
\n where $f(u)$ and $g(u)$ are arbitrary functions and $k$ is a positive 
constant less than $1$. The physical parameters would read as
\bq
 m(u,r) = f(u) - \frac {g(u)} {(kn - 1) r^{kn-1}}\cr
8 \pi \mu = -\frac {n} {(n - 1) r^n} \left( \dot f - \frac {\dot g} {(kn -1) r^{kn -1}} \right) \cr
8 \pi \rho = \frac {ng(u)} {(n-1) r^{n(k-1}}, p = k\rho \label{32}
\eq

 \n If $g(u)$ is positive, obviously $\rho\geq 0$ and $\rho\geq p$. 
Similar to the previous case the main restriction would come from 
$\mu\geq 0$. This would mean either (i) $\dot f< 0, \dot g > g 0$ and $k< g 
1/n$ or (ii) $\dot f< 0, \dot g < 0$ and $k/g >/n$.

 When $k = 1$, the above solution reduces to Bonnor-Vaidya solution 
discussed in the previous case. The metric explicitly reads as
\bq
 ds^2 =2 dudr + \left[ 1 - \frac {2f(u)} {(n-1) r^{n-1}} + {2g(u)} {(n-1) (kn -1)r^{n(k+1)-2}}\right] du^2 \label {33}
\eq

 \n This is Husain solution [17] in higher dimensions. It is 
asymptotically flat for $k < 1/n$ representing a bounded source while it 
is cosmological for $k > 1/n$. When $kn = 1$, $m(u,r) = f(u) + g(u) 
ln r$, it can be seen that the enrgy conditons are always violated for 
sufficiently small $r$ and hence this case is ruled out. We shall therefore 
consider the other two cases, one for bounded source and other for 
cosmological model.

 Case $k > 1/n$. For simplicity let us first set $k = 1$. Then we write 
$2f = A(1 - tanh u)$ and $2g = 1 + B tanh u$ with the constants $A \geq 0, 
0 \leq B \leq 1$, then all the energy conditions are satisfied. The metric 
then takes the form
\bq
 ds^2 = 2dudr + \left[ 1 - \frac {A(1- \tan h u)} {(n-1) r^{n-1}} + \frac {1 +B \tan h u} {(n-1)^2 r^{2(n-1}}\right] du^2\label{34}
\eq

\n In the limit $u \rightarrow  \infty$, it has a naked singularity at $r = 0$. 
However for $u \rightarrow  -\infty$, it may have horizons depending upon the 
realtive values of $A$ and $B$. Specifically horizons are given by
\bq
 (n-1) r^{n-1} = A \pm \sqrt {A^2 + B-1} \label{35}
\eq

 \n In the other case $k < \frac {1} {n} $, we write $2f = C + A(1 - tanh u), 2g = B(1 
- tan h u)$ satisfying the energy conditions. The metric 
\bq
 ds^2 = 2dudr + \left[ 1 - \frac {A(1- \tan h u)} {(n-1) r^{n-1}} +\frac {B (1 - \tan h u} {(n-1)(kn-1)r^{n(k+1)-2}}\right] du^2 \label{36}
\eq
\n where $A,B \geq 0$. In the limit $u \rightarrow  \infty$, the metric will 
either have naked singularity at $r = 0$ for $C \neq 0$ or it would be 
flat for $C = 0$. On the other side as $u \rightarrow  -\infty$ with $C = 0$, 
there would occur apparent horizons given by
\bq
 R^{\frac {n(k+1)-2} {n-1}} - \frac {2A} {n-1} R^{\frac {nk-1} {n-1}} + \frac {2B} {(n-1)(nk-1)} = 0, R= r^{n-1} \label{37}
\eq
\n In particular when $n =2$ and $k = 1/3$, the above equation takes the 
form
\bq
 (r - 2A)^3 = (3B)^3 r \label{38}
\eq
\n which would always have a real root for permissible values of $A$ and 
$B$.

\section{Discussion}

 We see that the 4-dimensional spherically symmetric solutions describing 
Type II fluid go over to $(n+2)$-dimensional spherically symmetric 
solutions and essentially retaining their physical behaviour. In 
particular higher dimensional version of Husain solution [17] that 
describes gravitational collapse leading to asymptotically flat black 
hole solutions for $k > 1/n$. The general metric depends upon the 
parameter $k$ and two arbitrary functions of retarded coordinate $u$, 
which are constrained by the enrgy conditions. Also the long retarded 
time limit of the asymptotically flat solutions would fall between 
Schwarzschild and Reissner-Nordetr$\ddot o$m solutions as in [17] in the 
sense that $1/n l k leq 1$ in (ref{31,32}). 

 In general the $k/g 1/n$ solutions represent evolution of naked 
singularity into itself or into black holes and the constant parameters 
in (ref{33}) determine which of these two possibilities would occur. In 
the other case of $k\l 1/n$, it is evolution of naked singularity or flat 
space  into black hole in a cosmological background. Of course in all the 
cases the cosmic censorship is respected. As in [17], we also have 
$T_{ik} l^il^k = 0$ even though the distribution is along both the null 
directions.

 It would however be possible to find more exact solutions of the similar 
kinds by imposing the equation of state $p = k\rho$. The main reason for 
clubbing together of so many cases is that physical parameters involve only 
linear derivatives of the mass function $m(u,r)$. Thus linear 
combinations of all the cases discussed above would also be a solution. 
For example if we take the equation of state $p = \rho - 
\bigwedge/{4\pi}$, the resulting solution is the higher dimensional 
Bonnor-Vaidya metric with the cosmological constant, which is a linear 
combination of the cases (a) and (d). Also note that it is trivially 
possible to include a global monopole or string dust in all the above 
cases. It turns out that solutions with global monopole could be 
considered as dual to the Einstein solutions in a sense that they 
are solutions of dual equation [24]. The dual solutions simply imbibe global 
monopole or string dust in the original spacetime, and they are generally 
not asymptotically flat.

Acknowledgement: LKP thanks IUCAA for hospitality.

\end{document}